\documentstyle[pra,aps]{revtex}
\newcommand{\beq}{\begin{equation}}
\newcommand{\eeq}{\end{equation}}
\newcommand{\bea}{\begin{eqnarray}}
\newcommand{\eea}{\end{eqnarray}}

\draft
\begin{document}
\title{Interference-induced gain in Autler-Townes doublet of a V-type atom in a
cavity}
\author{Peng Zhou$^{1,2}$\thanks{
Electronic address: peng.zhou@physics.gatech.edu}, S. Swain$^2$,
and L. You$^1$}
\address{$^1$School of Physics, Georgia Institute of Technology,
Atlanta, GA 30332-0430,\\$^2$Department of Applied Mathematics and
Theoretical Physics, \\ The Queen's University of Belfast, Belfast
BT7 1NN, the United Kingdom. }
\date{September, 1999}
\maketitle

\begin{abstract}
We study the Autler-Townes spectrum of a V-type atom coupled to a
single-mode, frequency-tunable cavity field at finite termperature, with a
pre-selected polarization in the bad cavity limit, and show that, when the
mean number of thermal photons $N\gg 1$ and the excited sublevel splitting
is very large (the same order as the cavity linewidth), the probe gain may
occur at either sideband of the doublet, depending on the cavity frequency,
due to the cavity-induced interference.
\end{abstract}


Within recent years, there has been a resurgence of interest in the
phenomenon of quantum interference \cite{arim96}. The principal reason is
that it lies at the heart of many new effects and applications of quantum
optics, such as lasing without population inversion \cite{harris},
electromagnetically-induced transparency \cite{EIT}, enhancement of the
index of refraction without absorption \cite{index}, fluorescence quenching
\cite{cardimona,zhou1,zhu1,zhu2} and spectral line narrowing \cite{zhou1}.

The basic system consists of a singlet state connected to a closely-spaced
excited doublet by a single-mode laser. Cardimona {\em et al.} \cite
{cardimona,zhou1} studied the effect of quantum interference on the
resonance fluorescence of such a system, and found that it can be driven
into a dark state in which quantum interference prevents any fluorescence
from the excited sublevels, regardless of the intensity of the exciting
laser. We have recently shown that quantum interference can also lead to
narrow resonances, transparency and gain without population inversion in the
probe absorption spectrum of such an atomic system \cite{zhou2}.

Harris and co-workers \cite{harris} generalized the V-type atom to systems
where the excited doublets decay to an additional continuum or to a single
auxiliary level, in addition to the ground state. They found that at a
certain frequency, the absorption rate goes to zero due to destructive
interference whereas the emission rate remains finite. It is possible to
amplify a laser field at this frequency without population inversion being
present. In the case of a single auxiliary level, quantum interference can
lead to the elimination of the spectral line at the driving laser frequency
in the spontaneous emission spectrum \cite{zhu1} and transparency in the
absorption spectrum \cite{plk}.

It is important for these effects that the dipole moments of the transitions
involved are parallel, so that the {\em cross-decay terms} are maximal. From
the experimental point view, however, it is difficult to find isolated
atomic systems which have parallel moments \cite
{harris,cardimona,agarwal,berman}.

Various alternative proposals \cite{agarwal,scully} have been made for
generating quantum interference effects. For example, if the two upper
levels of a V-type atom are coupled by a microwave field or an applied
laser, the excited doublet becomes a superposition, so that as the atom
decays from one of the excited sublevels it drives the other. For such
systems, the cross-decay terms are evident in the atomic dressed picture
\cite{scully}. A four-level atom with two closely-spaced intermediate states
coupled to a two-mode cavity can also show the effect of quantum
interference \cite{agarwal}. In fact, the experimental observation of the
interference-induced suppression of spontaneous emission was carried out in
sodium dimers where the excited sublevels are superpositions of singlet and
triplet states that are mixed by a spin-orbit interaction \cite{zhu2,berman}.

We have recently also proposed a scheme for engineering of quantum
interference (parallel or anti-parallel dipole moments) in a
V-type atom coupled to a frequency tunable, single-mode cavity
field with a pre-selected polarization at zero temperature
\cite{zhou3}. We have found that the effects of the
cavity-induced interference are pronounced only when the cavity
detuning $\delta $ and the excited doublet splitting $\omega
_{21}$ are much less than the cavity linewidth $2\kappa $. Here
we shall extend the study to a cavity damped by {\it a thermal
reservoir} at finite temperature, so that the mean number of
thermal photons, $N$, in the cavity mode is nonzero. We show
that, even in the case of $\delta $ and $\omega _{21}$ being the
same order of the cavity linewidth $2\kappa $, the cavity-induced
interference is still significant when $N\gg 1$, and that
interference-assisted gain may occur in one component of the
Autler-Townes doublet for certain cavity resonant frequency. Such
interference-related gain in the Autler-Townes doublet is also
reported in free space \cite{mg}.

Our model consists of a V-type atom with the ground state
$|0\rangle $ coupled by the single-mode cavity field to the
excited doublet $|1\rangle ,\,|2\rangle $. Direct transitions
between the excited sublevels $|1\rangle $ and $|2\rangle $ are
dipole forbidden. The master equation for the total density
matrix operator $\rho _{T}$ in the frame rotating with the
average atomic transition frequency $\omega _{0}=(\omega
_{10}+\omega _{20})/2$ takes the form

\begin{equation}
\dot{\rho}_{T}=-i\left[ H_{A}+H_{C}+H_{I},\,\rho _{T}\right] +{\cal L}\rho
_{T},  \label{master1}
\end{equation}
where
\begin{mathletters}
\begin{eqnarray}
H_{C} &=&\delta \,a^{\dagger }a, \\
H_{A} &=&\frac{1}{2}\omega _{21}\left( A_{22}-A_{11}\right) , \\
H_{I} &=&i\left( g_{1}A_{01}+g_{2}A_{02}\right) a^{\dagger }-h.c., \\
{\cal L}\rho _{T} &=&\kappa (N+1)\left( 2a\rho _{T}a^{\dagger }-a^{\dagger
}a\rho _{T}-\rho _{T}a^{\dagger }a\right) \\
&&+\kappa N\left( 2a^{\dagger }\rho _{T}a-aa^{\dagger }\rho _{T}-\rho
_{T}aa^{\dagger }\right) ,
\end{eqnarray}
with
\end{mathletters}
\begin{equation}
\delta =\omega _{C}-\omega _{0},\;\;\;\;\omega
_{21}=E_{2}-E_{1},\;\;\;\;g_{i}={\bf e}_{\lambda }\cdot {\bf d}_{0i}\sqrt{%
\frac{\hbar \omega _{C}}{2\epsilon _{0}V}},\;\;\;\;(i=1,\,2).
\end{equation}
Here $H_{C}$, $H_{A}$ and $H_{I}$ are the unperturbed cavity, the
unperturbed atom and the cavity-atom interaction Hamiltonians respectively,
while ${\cal L}\rho _{T}$ describes damping of the cavity field by the
continuum electromagnetic modes at finite temperature, characterized by the
decay constant $\kappa $ and the mean number of thermal photons $N$; $a$ and
$a^{\dag }$ are the photon annihilation and creation operators of the cavity
mode, and $A_{ij}=|i\rangle \langle j|$ is the atomic population (the dipole
transition) operator for $i=j$ $(i\neq j)$; $\delta $ is the cavity detuning
from the average atomic transition frequency, $\omega _{21}$ is the
splitting of the excited doublet of the atom, and $g_{i}$ is the atom-cavity
coupling constant, expressed in terms of ${\bf d}_{ij},$ the dipole moment
of the atomic transition from $|j\rangle $ to $|i\rangle ,$ ${\bf e}%
_{\lambda }$, the polarization of the cavity mode, and $V,$ the volume of
the system. In the remainder of this work we assume that the polarization of
the cavity field is {\em pre-selected}, i.e., the polarization index $%
\lambda $ is fixed to one of two possible directions.

In this paper we are interested in the bad cavity limit: $\kappa \gg g_{i}$,
that is the atom-cavity coupling is weak, and the cavity has a low $Q$ so
that the cavity field decay dominates. The cavity field response to the
continuum modes is much faster than that produced by its interaction with
the atom, so that the atom always experiences the cavity mode in the state
induced by the thermal reservoir. Thus one can adiabatically eliminate the
cavity-mode variables, giving rise to a master equation for the atomic
variables only \cite{detail}, which takes the form,

\begin{eqnarray}
\dot{\rho} &=&-i\left[ H_{A},\;\rho \right]  \nonumber \\
&&+\{F(\omega _{21})(N+1)\left[ |g_{1}|^{2}\left( A_{01}\rho
A_{10}-A_{11}\rho \right) +g_{1}g_{2}^{*}\left( A_{01}\rho A_{20}-A_{21}\rho
\right) \right]  \nonumber \\
&&+F(-\omega _{21})(N+1)\left[ |g_{2}|^{2}\left( A_{02}\rho
A_{20}-A_{22}\rho \right) +g_{1}^{*}g_{2}\left( A_{02}\rho A_{10}-A_{12}\rho
\right) \right]  \nonumber \\
&&+F(\omega _{21})N\left[ |g_{1}|^{2}\left( A_{10}\rho A_{01}-\rho
A_{00}\right) +g_{1}g_{2}^{*}A_{20}\rho A_{01}\right]  \nonumber \\
&&+F(-\omega _{21})N\left[ |g_{2}|^{2}\left( A_{20}\rho A_{02}-\rho
A_{00}\right) +g_{1}^{*}g_{2}A_{10}\rho A_{02}\right]  \nonumber \\
&&+h.c.\}  \label{master}
\end{eqnarray}
where $F(\pm \omega _{21})=\left[ \kappa +i(\delta \pm \omega
_{21}/2)\right] ^{-1}$.

Obviously, the equation (\ref{master}) describes the cavity-induced atomic
decay into the cavity mode. The real part of $F(\pm \omega _{21})|g_{j}|^{2}$
represents the cavity-induced decay rate of the atomic excited level $%
j\,(=1,\,2)$, while the imaginary part is associated with the frequency
shift of the atomic level resulting from the interaction with the vacuum
field in the detuned cavity. The other terms, $F(\pm \omega
_{21})g_{i}g_{j}^{*},\,(i\neq j)$, however, represent the cavity-induced
correlated transitions of the atom, {\it i.e.}, an emission followed by an
absorption of the same photon on a different transition, ($|1\rangle
\rightarrow |0\rangle \rightarrow |2\rangle $ or $|2\rangle \rightarrow
|0\rangle \rightarrow |1\rangle $), which give rise to the effect of quantum
interference.

The effect of quantum interference is very sensitive to the orientations of
the atomic dipoles and the polarization of the cavity mode. For instance, if
the cavity-field polarization is not pre-selected, as in free space, one
must replace $g_{i}g_{j}^{*}$ by the sum over the two possible polarization
directions, giving $\Sigma _{\lambda }g_{i}g_{j}^{*}\propto {\bf d}%
_{0i}\cdot {\bf d}_{0j}^{*}$ \cite{agarwal}. Therefore, only non-orthogonal
dipole transitions lead to nonzero contributions, and the maximal
interference effect occurs with the two dipoles parallel. As pointed out in
Refs. \cite{harris,cardimona,agarwal,berman} however, it is questionable
whether there is a isolated atomic system with parallel dipoles. Otherwise,
if the polarization of the cavity mode is fixed, say ${\bf e}_{\lambda }=%
{\bf e}_{x}$, the polarization direction along the $x$-quantization axis,
then $g_{i}g_{j}^{*}\propto \left( {\bf d}_{0i}\right) _{x}\left( {\bf d}%
_{0j}^{*}\right) _{x}$, which is nonvanishing, regardless of the orientation
of the atomic dipole matrix elements.

It is apparent that if $\kappa \gg \delta ,\,\omega _{21}$, the frequency
shifts are negligibly small \cite{zhou3} , and this equation (\ref{master})
reduces to that of a V-atom with two parallel transition matrix elements in
free space \cite{cardimona,zhou1,zhou2}. In the following we shall discuss
the effect of quantum interference in the situation of $\omega_{21} \geq
\kappa$ and $N\gg 1$, by examining the steady-state absorption spectrum of
such a system, which is defined as

\begin{equation}
A(\omega )=\Re e\int_{0}^{\infty }\lim_{t\rightarrow \infty }\left\langle
\left[ P(t+\tau ),\;P^{\dag }(t)\right] \right\rangle e^{i\omega \tau }d\tau
,
\end{equation}
where $\omega =\omega _{p}-\omega _{0}$, and $\omega _{p}$ is the frequency
of the probe field and $P(t)=d_{1}A_{01}+d_{2}A_{02}$ is the component of
the atomic polarization operator in the direction of the probe field
polarization vector ${\bf e}_{p}$, with $d_{i}={\bf e}_{p}\cdot {\bf d}_{0i}$%
. With the help of the quantum regression theorem, one can calculate the
spectrum from the Bloch equations,

\begin{eqnarray}
\langle \dot{A}_{11}\rangle &=&-\left[ F(\omega _{21})+F^{*}(\omega
_{21})\right] |g_{1}|^{2}\left[ (N+1)\langle A_{11}\rangle -N\langle
A_{00}\rangle \right]  \nonumber \\
&&-F(-\omega _{21})g_{1}^{*}g_{2}(N+1)\langle A_{12}\rangle -F^{*}(-\omega
_{21})g_{1}g_{2}^{*}(N+1)\langle A_{21}\rangle ,  \nonumber \\
\langle \dot{A}_{22}\rangle &=&-\left[ F(-\omega _{21})+F^{*}(-\omega
_{21})\right] |g_{2}|^{2}\left[ (N+1)\langle A_{22}\rangle -N\langle
A_{00}\rangle \right]  \nonumber \\
&&-F^{*}(\omega _{21})g_{1}^{*}g_{2}(N+1)\langle A_{12}\rangle -F(\omega
_{21})g_{1}g_{2}^{*}(N+1)\langle A_{21}\rangle ,  \nonumber \\
\langle \dot{A}_{12}\rangle &=&-F(\omega _{21})g_{1}g_{2}^{*}(N+1)\langle
A_{11}\rangle -F^{*}(-\omega _{21})g_{1}g_{2}^{*}(N+1)\langle A_{22}\rangle
+\left[ F(\omega _{21})+F^{*}(-\omega _{21})\right] g_{1}g_{2}^{*}N\langle
A_{00}\rangle  \nonumber \\
&&-\left[ F^{*}(\omega _{21})|g_{1}|^{2}(N+1)+F(-\omega
_{21})|g_{2}|^{2}(N+1)+i\omega _{21}\right] \langle A_{12}\rangle ,
\nonumber \\
\langle \dot{A}_{01}\rangle &=&-\left[ F(\omega
_{21})|g_{1}|^{2}(2N+1)+F(-\omega _{21})|g_{2}|^{2}N-i\frac{\omega _{21}}{2}%
\right] \langle A_{01}\rangle -F(-\omega _{21})g_{1}^{*}g_{2}(N+1)\langle
A_{02}\rangle ,  \nonumber \\
\langle \dot{A}_{02}\rangle &=&-\left[ F(\omega _{21})|g_{1}|^{2}N+F(-\omega
_{21})|g_{2}|^{2}(2N+1)+i\frac{\omega _{21}}{2}\right] \langle A_{02}\rangle
-F(\omega _{21})g_{1}g_{2}^{*}(N+1)\langle A_{01}\rangle .
\end{eqnarray}

To monitor quantum interference, we insert a factor $\eta \,(=0,\,1)$ in the
cross transition terms $g_{i}g_{j}^{*}$. When $\eta =0$, the cross
transitions are switched off, so no quantum interference is present.
Otherwise, the effect of quantum interference is maximal.

Figure 1 shows the Autler-Townes spectra for $g_{1}=g_{2}=10$, $\kappa
=\omega _{21}=100,\,N=10$, and different cavity detunings. In the absence of
the interference ($\eta =0$),{\it \ } two transition paths, $|0\rangle
\leftrightarrow |1\rangle $ and $|0\rangle \leftrightarrow |2\rangle $, are
{\it independent}, which{\it \ }lead to the lower and higher frequency
sidebands of the absorption doublet, respectively. It is not difficult to
see that the spectral heights and linewidths are mainly determined by the
cavity-induced decay constants $\gamma _{i}\,(i=1,\,2)$ of the excited
states, which have the forms
\begin{equation}
\gamma _{1}=\frac{\kappa |g_{1}|^{2}}{\kappa ^{2}+(\delta +\omega
_{21}/2)^{2}},\;\;\;\gamma _{2}=\frac{\kappa |g_{2}|^{2}}{\kappa
^{2}+(\delta -\omega _{21}/2)^{2}},  \label{gam}
\end{equation}
which vary with the cavity frequency. It is evident that $\gamma _{1}<\gamma
_{2}$ when $\delta >0$, and both $\gamma _{1}$ and $\gamma _{2}$ decrease as
$\delta $ increases. Noting that the lower and higher frequency peaks have
respective liewidths $\Gamma _{l}=\gamma _{1}(2N+1)+\gamma _{2}N$ and $%
\Gamma _{h}=\gamma _{1}N+\gamma _{2}(2N+1)$, and are proportional to $\Gamma
_{l,h}^{-1}$, the lower frequency sideband is slightly higher than the
higher frequency one in the case of $\delta >0$ and both the sidebands can
be narrowed by increasing the cavity detuning, see for example, the dashed
lines in the following three figures.

Whereas, the spectral features are dramatically modified in the presence of
the cavity induced interference ($\eta =1$). When the cavity is resonant
with the average frequency of the atomic transitions, $\delta =0$, the
doublet is symmetric, and its sidebands are higher and wider than that for $%
\eta =0$, as shown in the frames 1(a), 2(a) and 3(a). Otherwise, it is
asymmetric. Either sideband of the doublet can be suppressed, depending upon
the cavity frequency, {\it e.g.}, the higher frequency sideband is
suppressed for $\delta =10$, $50$ and $100$, see in Figs. 1(b)--1(d), while
the sideband is enhanced for $\delta =200$, shown in Fig. 1(e). When the
cavity frequency is far off resonant with the atomic transition frequencies,
say $\delta =500$ in Fig. 1(f), the absorption spectra for $\eta =0$ and $1$
are virtually same, that is the effect of the cavity induced interference is
negligible small.

Rather surprisingly, the frame 1(c) shows probe gain in the higher frequency
sideband, without the help of any coherent pumping. Moreover, increasing the
mean number of thermal photons $N$ may enhance the probe gain, see for
instance, in Fig. 2 for $N=20$, in which the higher-frequency probe gain
even occurs for a relative small cavity detuning, say $\delta =10$ in the
frame 2(b). Contrastively, when the detuning is very large, the probe beam
can be amplified at the lower-frequency sideband, rather than at the
higher-frequency one, as shown in the frame 2(e) for $\delta =200$ for
example. Fig. 2 also exhibits that the linewidths are broadened for large
number of thermal photons.

We present the Autler-Townes spectrum for a large excited level-splitting, $%
\omega _{21}=200$, and a large number of thermal photons, $N=20$, in Fig. 3,
in which the more pronounced gain, comparing with that for $\omega _{21}=100$%
, is displayed at either the lower-frequency sideband for $\delta =10$, $50$
and $100$, or the higher-frequency sideband for $\delta =200$. One can also
find that for the large level-splitting, the effect of the cavity-induced
interference is still significant when $\delta =500$, as shown in Fig. 3(f),
where the lower frequency peak is almost suppressed while the other is
greatly enhanced. However, when $\delta \gg \omega _{21}$, say $\delta =1000$
for instance, the effect of the interference disappears (we have exhibited
no figure here).

In what follows, we shall see that the probe gain is a direct consequence of
the cavity-induced quantum interference between the two transition paths, $%
|0\rangle \leftrightarrow |1\rangle $ and $|0\rangle \leftrightarrow
|2\rangle $. The gain at different sidebands has different origin. To show
this, we first plot the steady-state population differences between the
excited sublevels and the ground level, $\langle A_{11}\rangle -\langle
A_{00}\rangle $ and $\langle A_{22}\rangle -\langle A_{00}\rangle $, and the
coherence between the excited sublevels, $\langle A_{12}\rangle $, against
the cavity detuning $\delta $ in Fig. 4 for $g_{1}=g_{2}=10$, $\kappa =100$,
$\omega _{21}=200$ and $N=20$. It is clearly that the steady-state
populations and coherence are highly dependent on the cavity frequency. The
coherence is symmetric with the cavity detuning and reaches the maximum
value at $\delta =0$, while the population differences are asymmetric.
Furthermore, the population inversion may be achieved for certain cavity
frequency, for example, if $143.8<\delta <650$, then $\langle A_{11}\rangle
-\langle A_{00}\rangle >0$, while $\langle A_{22}\rangle >\langle
A_{00}\rangle $ in the region of $-650<\delta <-143.8$. Therefore, the gain
in the region of $-143.8<\delta <143.8$ must stem from the cavity-induced
steady-state coherence between the two dipole-forbidden excited sublevels,
rather than from the population inversion between the two dipole transition
levels. Whereas, the population inversions may result in the probe gain when
the cavity detuning is in the regions of $-650<\delta <-143.8$ and $%
143.8<\delta <650$. We thus conclude that, in the case of $\delta >0,$ as
shown in Figs. 1-3, the gain at the lower-frequency sideband comes from the
contribution of the steady-state atomic coherence $\langle A_{12}\rangle $,
while the gain at the other sideband is attributed to the steady-state
population inversion $\left( \left\langle A_{11}\right\rangle >\left\langle
A_{00}\right\rangle \right) $.

Noting that, in the absence of the interference ($\eta =0$), $\langle
A_{11}\rangle =\langle A_{22}\rangle =N/(3N+1)$, $\langle A_{00}\rangle
=(N+1)/(3N+1)$, and $\langle A_{12}\rangle =0$ are independent of the cavity
detuning, the cavity frequency dependence of the steady-state populations
and coherence manifests the cavity-induced quantum interference.

To further explore the origin of the probe gain, we separate the
Autler-Townes spectrum into two parts, in which one corresponds to the
contribution of the populations, while the other results from the coherence,
in Fig. 5 for $g_{1}=g_{2}=10$, $\kappa =100$, $\omega _{21}=200$, $N=20$,
and various cavity frequencies. It is obvious that when $\delta =0$, $50$
and $100$, the contributions of the coherence to the spectrum are negative (%
{\it i.e.}, probe gain), whereas the populations make positive
contributions, see, for example, in frames 6(a)--6(c). One can also see that
the spectral component resulting from the populations is symmetric only when
$\delta =0$, otherwise, it has different values at the lower and higher
frequency sidebands, which are proportional to $\left( \left\langle
A_{00}\right\rangle -\left\langle A_{11}\right\rangle \right) $ and $\left(
\left\langle A_{00}\right\rangle -\left\langle A_{22}\right\rangle \right) $%
, respectively. As shown in Fig. 4, if the cavity detuning is zero, then $%
\left( \left\langle A_{00}\right\rangle -\left\langle A_{11}\right\rangle
\right) =\left( \left\langle A_{00}\right\rangle -\left\langle
A_{22}\right\rangle \right) $, whereas, $\left( \left\langle
A_{00}\right\rangle -\left\langle A_{11}\right\rangle \right) >\left(
\left\langle A_{00}\right\rangle -\left\langle A_{22}\right\rangle \right) $
for $\delta =50$ and $100$. As a result, the lower frequency peak is higher
than that of the other one in the cases of $\delta =50$ and $100$.
Therefore, the total spectrum may exhibit the probe gain at the higher
frequency sideband at these cavity frequencies, see, for example, in Figs.
3(c) and 3(d). The gain is purely attributed to the cavity-induced
steady-state atomic coherence. However, when $\delta =200$, the situation is
reverse: the coherence gives rise to the probe absorption, while the
populations lead to the gain at the lower frequency sideband, due to the
population inversion between the levels $|0\rangle $ and $|1\rangle $, as
illustrated in Fig. 4.

In summary, we have shown that maximal quantum interference can be achieved
in a V-type atom coupled to a single-mode, frequency-tunable cavity field at
finite temperature, with a pre-selected polarization in the bad cavity
limit. There are no special restrictions on the atomic dipole moments, as
long as the polarization of the cavity field is pre-selected. We have
investigated the cavity modification of the Autler-Townes spectrum of such a
system, and predicted the probe gain at either sideband of the doublet,
depending upon the cavity resonant frequency, when the excited sublevel
splitting is very large (the same order as the cavity linewidth) and the
mean number of thermal photons $N\gg 1$. The gain occurring at different
sidebands has the various origin: in the case of $\delta >0$, the lower
frequency gain is due to the nonzero steady-state coherence, while the
higher frequency one is attributed to the steady-state population inversion.
Both the nonzero coherence and population inversion originate from the
cavity-induced quantum interference.

\acknowledgments

This work is supported by the United Kingdom EPSRC. We gratefully
acknowledge conversations with G. S. Agarwal and Z. Ficek. We
would also like to thank S. Menon for bringing their paper
\cite{mg} to our attention.

\begin{figure}[tbp]
\caption{Absorption spectrum for $g_1=g_2=10$, $\kappa=100,\,
\omega_{21}=100, \, N=10$, and $\delta=0,\, 10,\, 50,\, 100, \, 200,\, 500$
in (a)--(f), respectively. In Figs. 1--3 the solid curves represent the
spectrum in the presence of the maximal interference ($\eta=1$), whilst the
dashed curves are the spectrum in the absence of the interference ($\eta=0$%
). }
\label{fig1}
\end{figure}

\begin{figure}[tbp]
\caption{Same as FIG. 1, but with $N=20$.}
\label{fig2}
\end{figure}

\begin{figure}[tbp]
\caption{Same as FIG. 1, but with $\omega=200$ and $N=20$.}
\label{fig3}
\end{figure}

\begin{figure}[tbp]
\caption{The steady-state population differences and coherence vs the cavity
detuning, for $g_1=g_2=10,\, \omega=200,\,N=20$ and $\eta=1$. The solid,
dashed and dot-dashed lines respectively represent $(\langle A_{11} \rangle
-\langle A_{00}\rangle)$, $(\langle A_{22} \rangle -\langle A_{00}\rangle)$
and Re($\langle A_{12} \rangle$).}
\label{fig4}
\end{figure}

\begin{figure}[tbp]
\caption{Different contributions to the absorption spectrum, for $%
g_1=g_2=10,\,\kappa=100,\, \omega_{21}=200, \, N=20, \, \eta=1$, and $%
\delta=0,\, 50,\, 100, \, 200$ in (a)--(d), respectively. The solid curves
represent the contributions of the population differences, whilst the dashed
curves are the ones of the coherences.}
\label{fig5}
\end{figure}

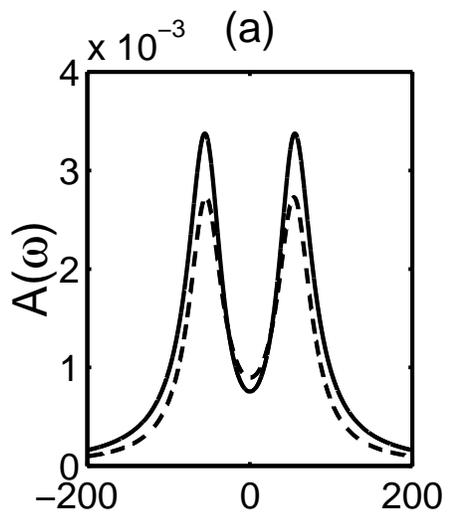
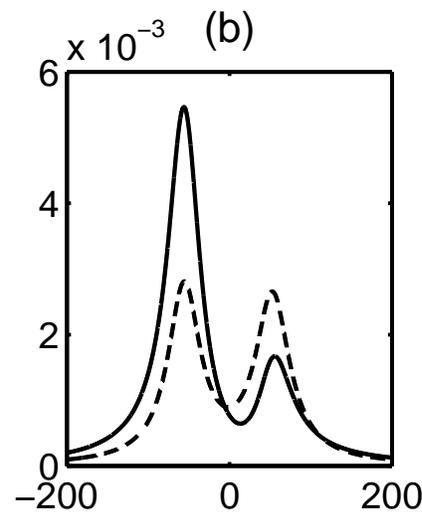
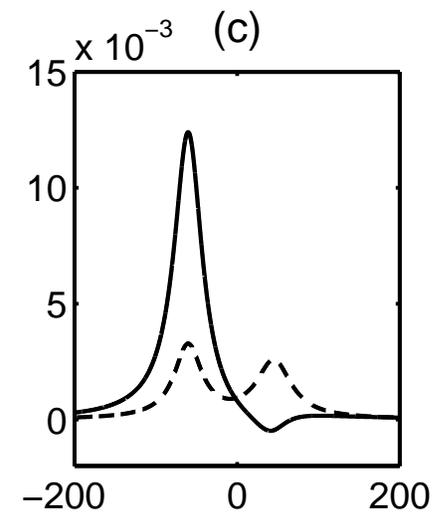
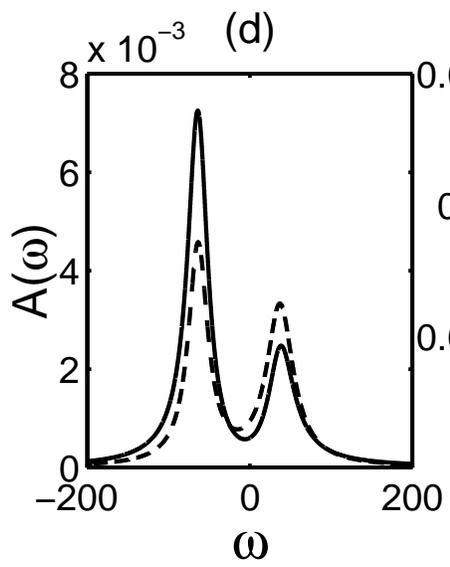
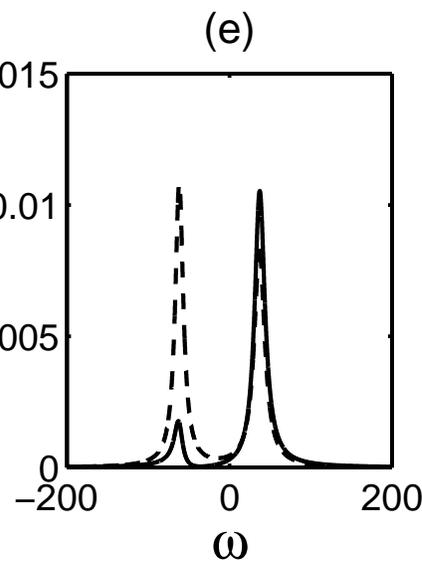
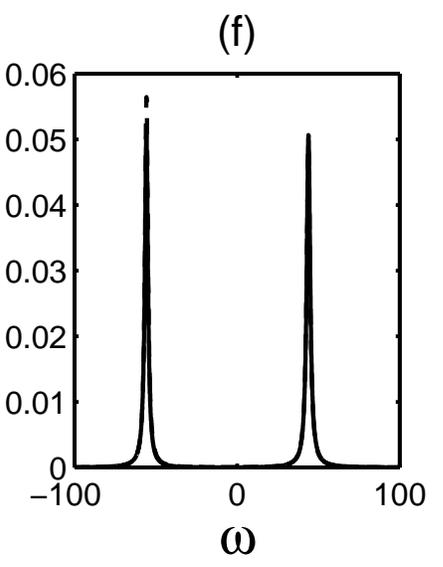
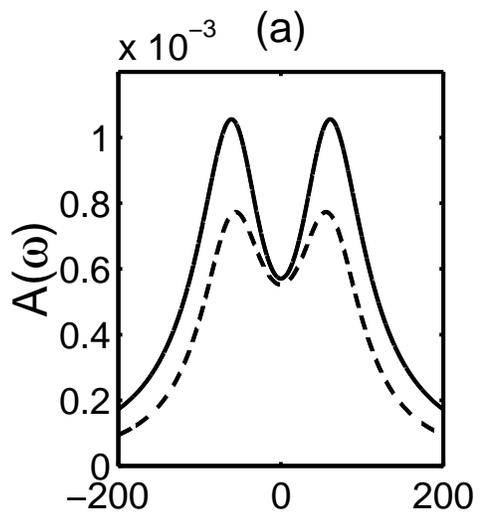
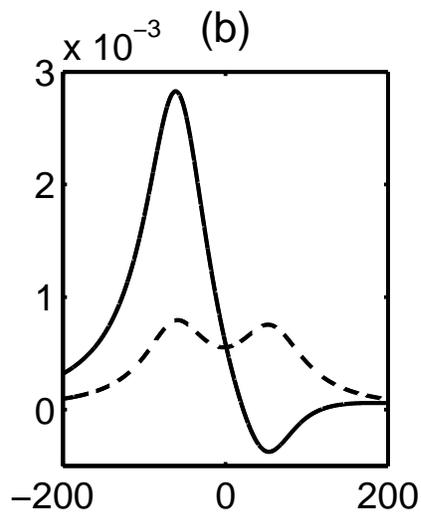
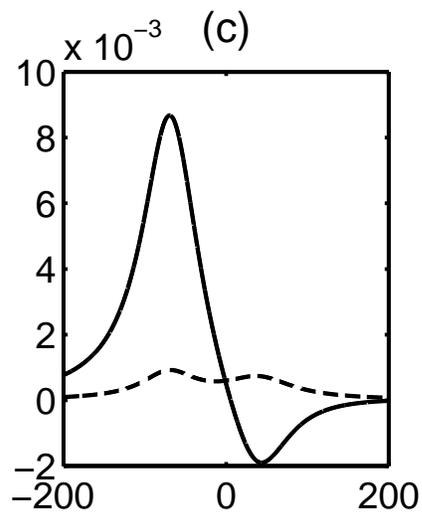
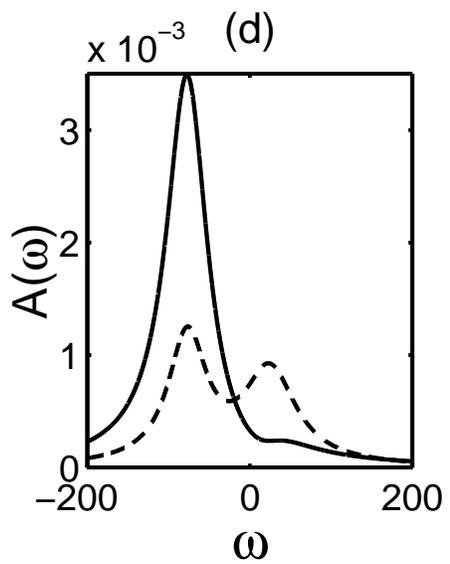
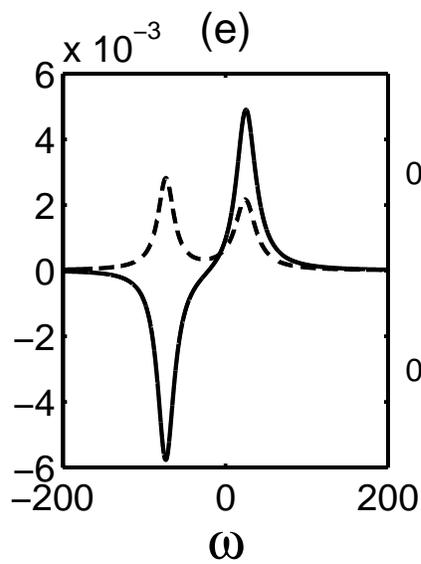
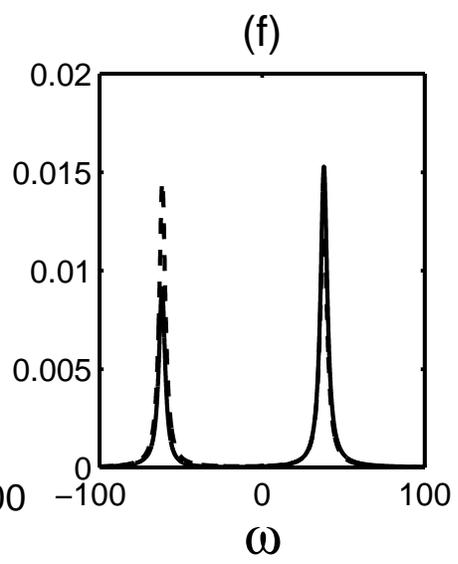
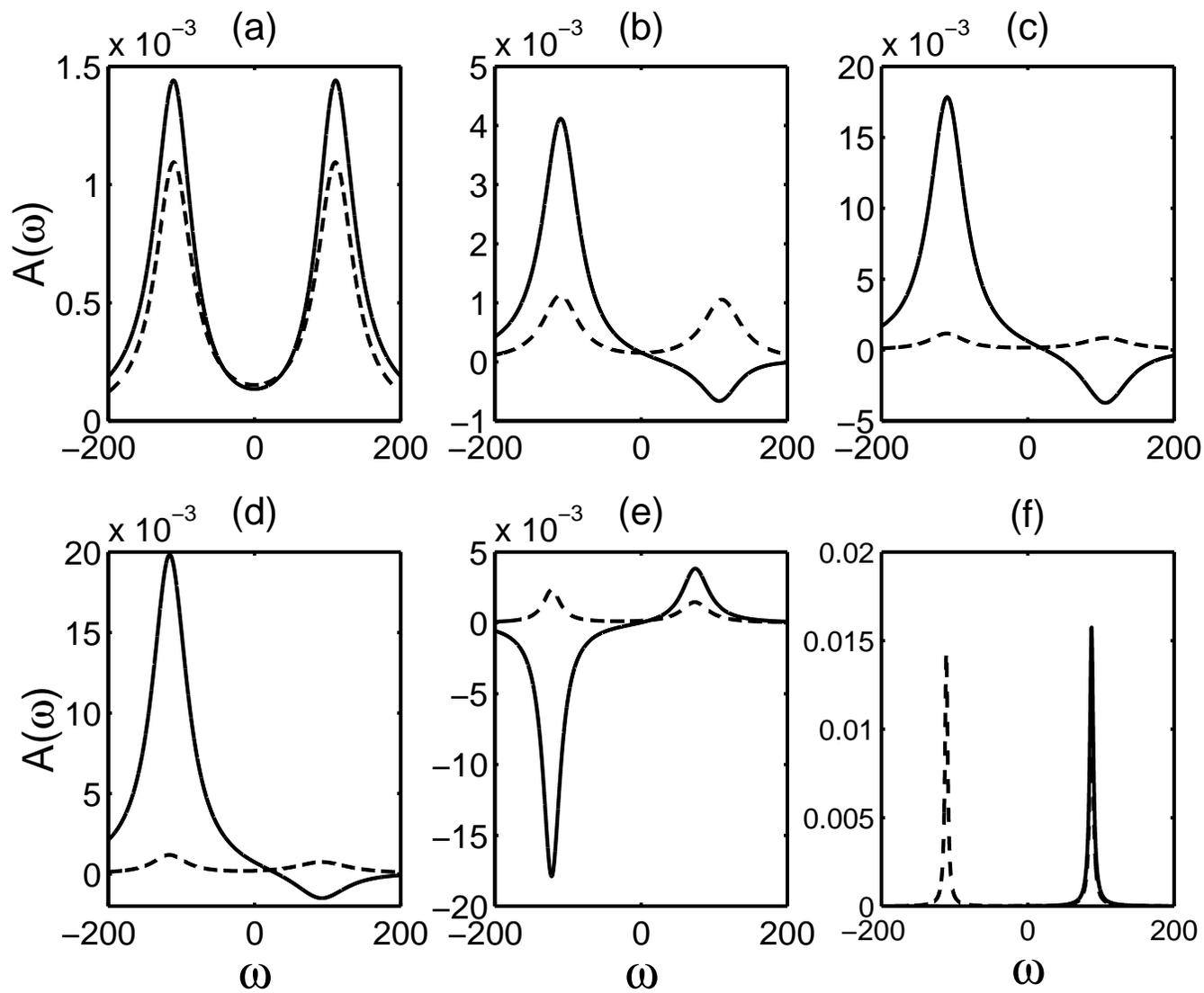
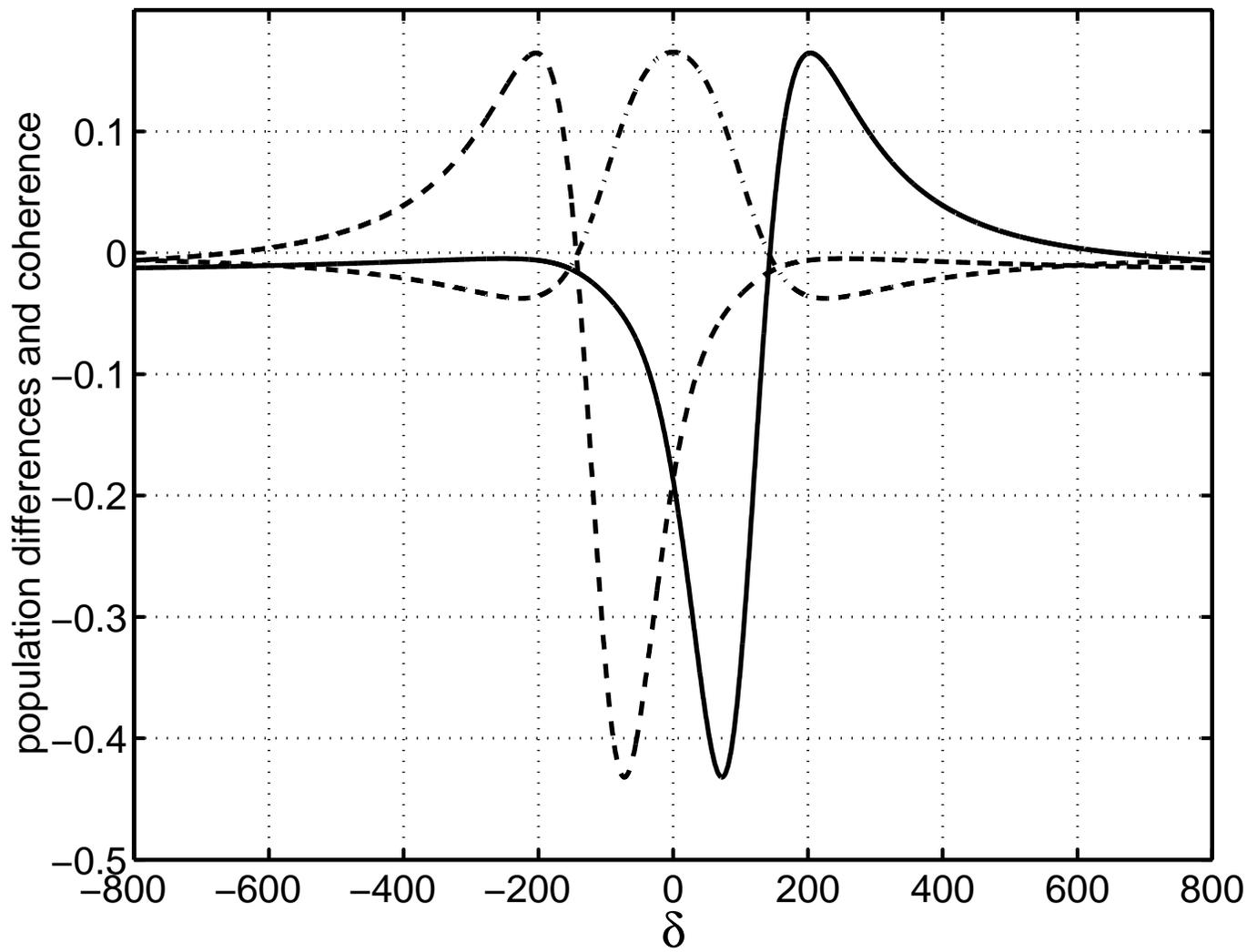
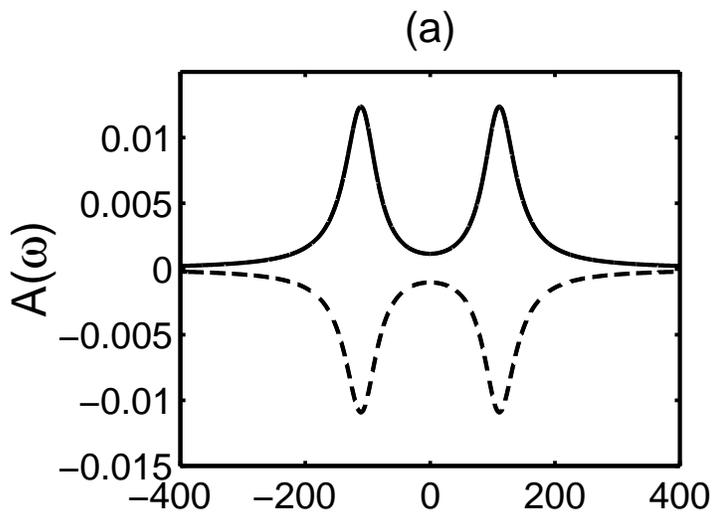
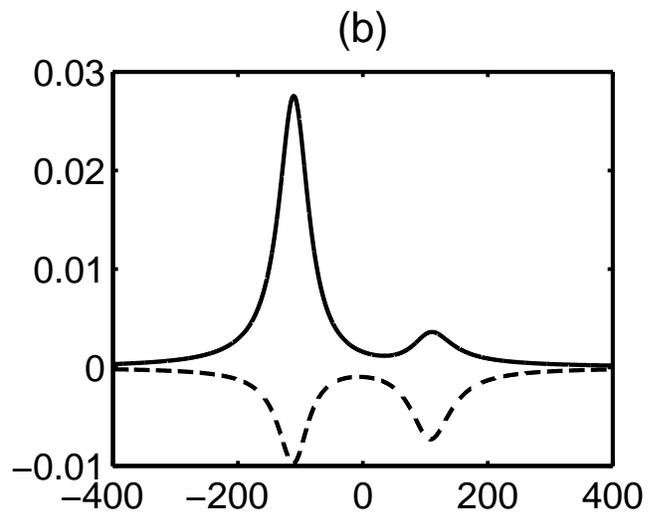
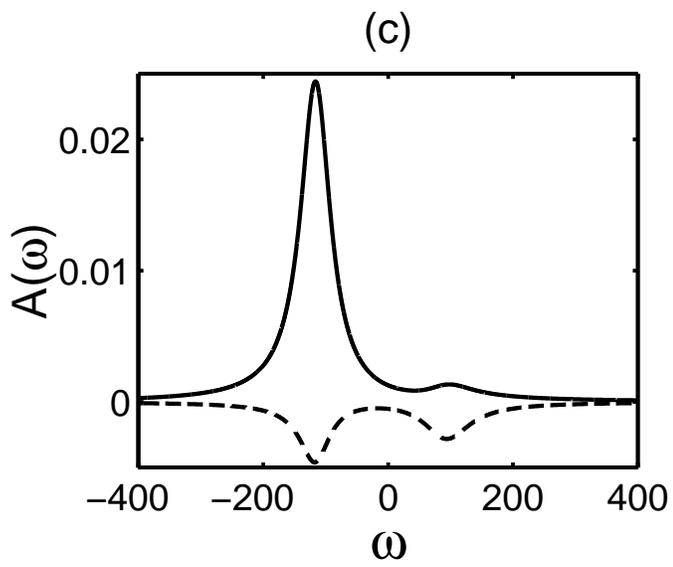
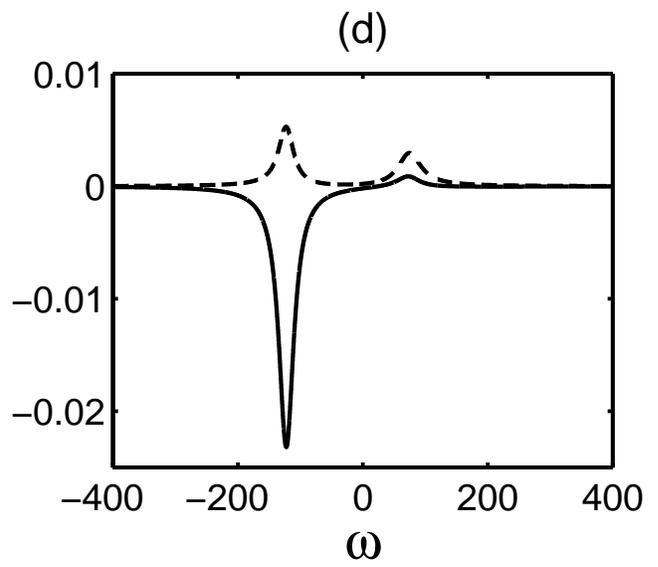
\end{document}